\documentclass{PoS}
\usepackage[latin1]{inputenc}
\usepackage{bm}
\usepackage{epsfig}
\usepackage{amsthm}
\usepackage{amsfonts}
\usepackage{float}
\usepackage{amsmath,amssymb}
\usepackage{graphicx}
\usepackage{dcolumn}
\usepackage{bm}
\def\nn{\nonumber}
\newcommand{\be}{\begin{equation}}
\newcommand{\ee}{\end{equation}}
\newcommand{\bea}{\begin{eqnarray}}
\newcommand{\eea}{\end{eqnarray}}

\newcommand{\om}{\omega}

\newcommand{\vk}{\vec k}

\newcommand{\vl}{\vec l}

\newcommand{\bpi}{\boldsymbol{\pi}}
\newcommand{\brho}{\boldsymbol{\rho}}

\newcommand{\del}{\partial}

\title{Shear viscosity of a pion gas due to $\rho\pi\pi$ 
and $\sigma\pi\pi$ interactions}

\ShortTitle{Shear viscosity of a pion gas}

\author{\speaker{Sabyasachi Ghosh}
\thanks{SG thanks to organizer of the Conference for giving the opportunity
to present this work as well as for supporting as student by waiving conference 
registration fees. 
Work partially financed by Funda\c{c}\~ao de Amparo \`a Pesquisa do Estado de 
S\~ao Paulo - FAPESP, Grant Nos. 2009/50180-0 (G.K.), 2012/16766-0 (S.G.), and
2013/01907-0 (G.K.); Conselho Nacional de Desenvolvimento Cient\'{\i}fico e 
Tecnol\'ogico - CNPq, Grant No. 305894/2009-9 (G.K.).}\\
Instituto de F\'{\i}sica Te\'orica, Universidade Estadual Paulista,
Rua Dr. Bento Teobaldo Ferraz, 271 - Bloco II, 01140-070 S\~ao Paulo, SP, Brazil.\\
        E-mail: \email{sabyaphy@gmail.com}}

\author{Gast\~ao Krein\\
        Instituto de F\'{\i}sica Te\'orica, Universidade Estadual Paulista,
Rua Dr. Bento Teobaldo Ferraz, 271 - Bloco II, 01140-070 S\~ao Paulo, SP, Brazil.\\
        E-mail: \email{gkrein@ift.unesp.br}}
        
\author{Sourav Sarkar\\
        Theoretical Physics Division, Variable Energy Cyclotron Centre, 
1/AF Bidhannagar, Kolkata 700064, India\\
        E-mail: \email{sourav@vecc.gov.in}}

\abstract{We have calculated the shear viscosity of pion medium, where
propagating pion has some finite thermal width due to interaction with the 
low mass resonances, $\sigma$ and $\rho$. With the help of standard
thermal field theoretical technique, the thermal width of the pion 
has been calculated from the pion self-energy diagram for
$\pi\sigma$ and $\pi\rho$ loops, where an effective Lagrangian density
has been used for interaction part. We have found
a very small value of shear viscosity by entropy density ratio ($\eta/s$),
which is very close to the KSS bound.}

\FullConference{7th International Conference on Physics and Astrophysics of Quark Gluon Plasma\\
		1-5 February , 2015\\
		Kolkata, India}

\begin{document}

\section{Introduction}
\label{sec:intro}
The new state of matter, which is expected to be produced 
due to high energy nucleus-nucleus collisions at Relativistic Heavy Ion Collider
(RHIC), is likely to have a very small ratio of shear 
viscosity to entropy density, $\eta/s$. This conclusion has been made by
different hydrodynamical and transport simulation to explain the elliptic 
flow parameter, $v_2$, extracted from data collected at RHIC.
Such a small $\eta/s$ is not really compatible with standard finite
temperature calculation of Quantum Chromo Dynamics (QCD), which exhibits
weakly interacting gas due to the asymptotic freedom of QCD at high temperature.
Therefore, several investigations on microscopic and model dependent 
calculation of $\eta/s$ for quark matter as well as hadronic matter have
been done in recent times. The latter one suddenly attract
some extra importance when Niemi et al.~\cite{Niemi} shows that the extracted 
transverse momentum $p_T$ dependence on elliptic flow parameter, $v_2(p_T)$, 
of RHIC data is highly sensitive to the temperature dependent $\eta/s$ 
in hadronic matter. In this context, we have calculated $\eta$ as well as $\eta/s$ 
of a pion gas using an effective Lagrangian for $\pi\pi\sigma$ and $\pi\pi\rho$
interactions and then we have also compared our results to others of the 
recent literature.   
\section{Formalism}
\label{sec:form}

Let us start with the standard expression of shear viscosity for pion gas,
obtained from relaxation time approximation (RTA)~\cite{Gavin,Prakash,SSS}: 
\be
\eta = \frac{\beta}{10\pi^2}\int\frac{d^3\vk\,\vk^6}{\Gamma_{\pi}(\vk,T)\,\om_k^2 } 
\, n(\om_k)\left[1+n(\om_k)\right] ,
\label{eta1_final}
\ee
where $n(\om_k) = 1/\{e^{\beta\om_k}-1\}$
is Bose-Einstein (BE) distribution function for a temperature $T = 1/\beta$, with 
$\om_k = (\vk^2 +m_\pi^2)^{1/2}$, and $\Gamma_\pi(\vk,T)$ is identified as the 
thermal width of $\pi$ mesons in the medium. 

As $\sigma$ and $\rho$ resonances come into the 
picture of $\pi$-$\pi$ scattering cross section,
therefore we have adopted an effective interaction Lagrangian density, 
\be
{\cal L} = g_\rho \, \brho_\mu \cdot \bpi \times \del^\mu \bpi 
+ \frac{g_\sigma}{2} m_\sigma \bpi\cdot\bpi\,\sigma,
\label{Lag}
\ee
as a dynamical guidance of pion interaction with mesonic medium. 
We have taken coupling constants $g_\rho=6$ and $g_\sigma=5.82$, which 
are fixed from experimental decay widths of $\rho$ and $\sigma$ respectively. 

\begin{figure}
\includegraphics[scale=0.55]{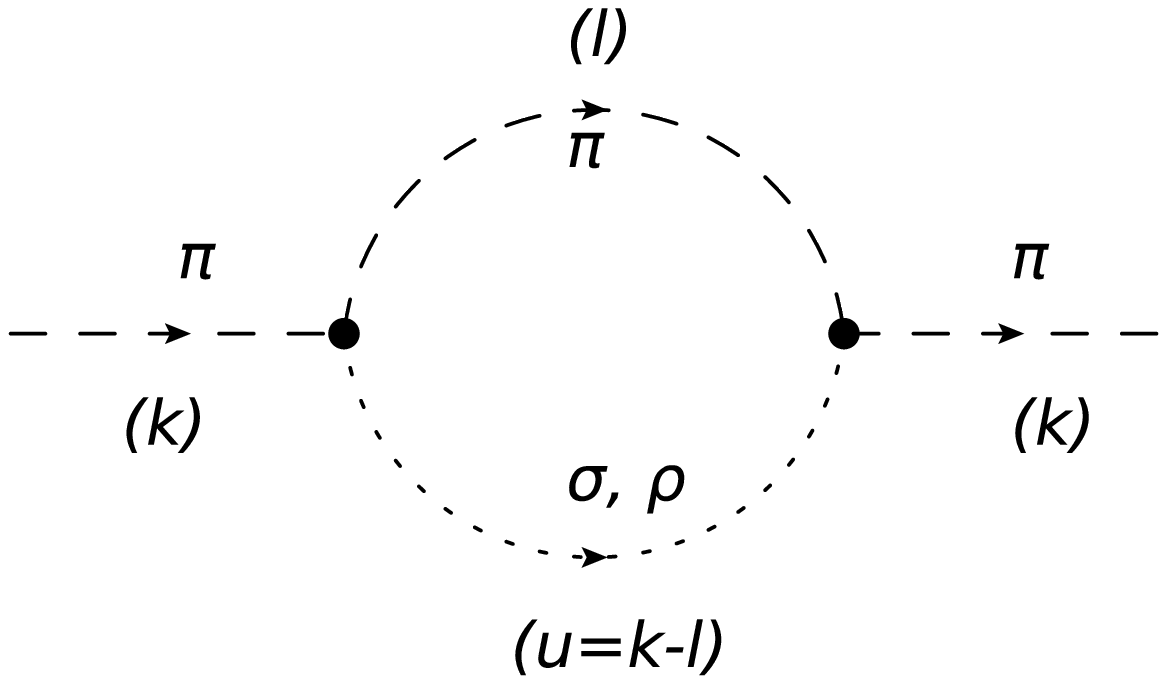}
\includegraphics[scale=0.35]{POS2.eps}
\caption{Left : General representation of pion self-energy diagram for $\pi\sigma$
and $\pi\rho$ loops.
Right : Momentum dependence of the thermal width
at temperature $T = 0.170$~GeV (upper panel) and $T = 0.150$~GeV (lower panel).} 
\label{Fig1_2}
\end{figure}

The pionic thermal width $\Gamma_{\pi}$ has been obtained from the
imaginary part of pion self-energy $\Pi_{\pi(\pi R)}$ for $\pi R$ loops,
where resonance $R$ stands for $\sigma$ and $\rho$ mesons. The self-energy
diagram is shown in the left panel of Fig.~(\ref{Fig1_2}).
Our required relation,
after thermal field theoretical deduction, is~\cite{GKS} 
\bea
\Gamma_{\pi}(\vk,T) &=& \sum_{R=\sigma, \rho}\Gamma_{\pi(\pi R)}
=\sum_{R=\sigma, \rho}\frac{{\rm Im}\Pi_{\pi(\pi R)}(k_0=\om_k,\vk, T)}{m_\pi}
\nn\\
&=&\sum_{R=\sigma, \rho}\frac{1}{m_\pi} \int \frac{d^3 \vl}{32\pi^2 \om_l\om_u}
L(k,l)|_{l_0=-\om_l,k_0=\om_k} \{n(\om_l) - n(\om_u)\}\delta(\om_k +\om_l - \om_u)~,
\label{G_pi_piM}
\eea
where $n(\om_l)$ and $n(\om_u)$ are BE distribution functions of
intermediate $\pi$ and $R$ mesons for $\om_l=\{\vl^2 + m_\pi^2\}^{1/2}$ 
and $\om_u=\{|\vk - \vl|^2 + m_R^2\}^{1/2}$ respectively. 
The vertex factors -
\bea
L(k,l) &=& - \frac{g^2_\sigma m_\sigma^2}{4}, 
~~~~~~~~~~~~~~~~~~~{\rm for}~R=\sigma~,
\nn\\
 &=& -\frac{g^2_\rho}{m_\rho^2} \, 
[ k^2 \left(k^2 - m^2_\rho\right) + 
l^2 \left(l^2 - m^2_\rho\right) 
- \, 2\{ (k\cdot l) \, m^2_\rho + k^2 \,l^2 \}],~{\rm for}~~~R=\rho
\eea
can be obtained from the effective Lagrangian density, given in Eq.~\ref{Lag}.

\vspace{0.5cm}
Next, to take into account the widths of the resonances, at first their
masses $m_R$ are taken as free parameter like invariant mass $M$. Then
the modified $\Gamma_{\pi(\pi R)}(\vk, T; M)$ of Eq.~(\ref{G_pi_piM})
are folded by vacuum spectral functions of corresponding resonances,
whose general form is
\be
\rho_R(M) = \frac{1}{\pi}{\rm Im}\left[\frac{1}{M^2-m_R^2+iM\Gamma_R(M)}\right]~.
\label{rho_u}
\ee
With the help of the Lagrangian density~\ref{Lag},
the vacuum decay width $\Gamma_R(M)$ of the $R=\sigma,\rho$ mesons
can be obtained as
\bea
\Gamma_\sigma (M) &=& \frac{3g_\sigma^2m_\sigma^2}{32\pi M} 
\left(1 - \frac{4m_\pi^2}{M^2}\right)^{1/2},
\label{s-width}
\\[0.25true cm]
\Gamma_\rho(M) &=& \frac{g_\rho^2M}{48\pi} 
\left(1 - \frac{4m_\pi^2}{M^2}\right)^{3/2}~.
\label{r-width} 
\eea
The normalized relation for this folding technique, which we have taken, is
\be
\Gamma_{\pi(\pi R)}(\vk,T,m_R) =  \frac{\int
dM^2  \, \rho_R(M) \, \Gamma_{\pi(\pi R)}(\vk,T;M)}
{\int dM^2  \, \rho_R(M)}~.
\label{gm_mu}
\ee
%
\section{Results and Discussion}
\label{sec:num}
\begin{figure}
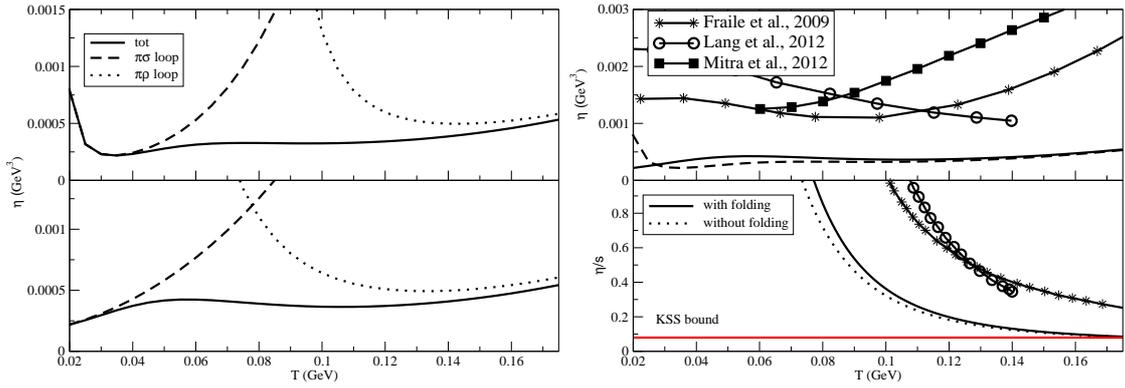

\includegraphics[scale=0.3]{eta_T.eps}
\includegraphics[scale=0.3]{eta_T_all.eps}
\caption{(Color online) Left : With (lower panel) and without (upper panel) folding 
results of $\eta$ vs $T$ due to $\pi\sigma$ (dashed lines), $\pi\rho$ (dotted lines) 
loops and their total (solid line). Right : Our estimation of $\eta(T)$ (upper panel) 
$\eta/s(T)$ (lower panel) have been compared to some other results. Horizontal red line
indicates the KSS bound of $\eta/s$.}
\label{fig:eta_T_rs}
\end{figure}

Using Eq.~(\ref{G_pi_piM}) and (\ref{gm_mu}), 
without (dotted line) and with (solid line) folding results of thermal width as 
a function of momentum are respectively estimated and 
displayed in the right panel of Fig.~\ref{Fig1_2} 
at $T = 0.170$~GeV (upper panel) and $T = 0.150$~GeV (lower panel). 
We see that due to folding effect, modification of $\Gamma_\pi(\vk)$ 
at low momentum becomes larger than that at high momentum and this
modification increases with the increasing of temperature. 
This momentum distribution of $\Gamma_\pi$ will be inversely integrated out
to obtain shear viscosity $\eta$ of pion gas, as expressed in Eq.~(\ref{eta1_final}).
The temperature dependence of $\eta$ is originated from two sources - the Bose-Einstein
distribution of pion and its thermal width $\Gamma_\pi$. 
In the left panel of Fig.~\ref{fig:eta_T_rs}, the temperature dependence 
of $\eta$ is presented using without (upper panel) and with (lower panel) 
folding results of $\Gamma_\pi$. 
The shear viscosity due to $\pi\rho$ (dotted line) 
and $\pi\sigma$ (dashed line) loops become
divergent in the higher ($T > 0.100$~GeV) and lower 
($T<0.100$ GeV) temperature regions respectively. This complementary
features of these two loops indicates that 
consideration of both resonances in $\pi-\pi$ scattering is strictly 
necessary to obtain a smooth, non divergent $\eta$ for temperatures below 
the critical, $T_c \simeq 0.175$~GeV. Though without folding results of $\eta$ 
at very low temperatures ($T<0.020$ GeV) tends to diverge, but 
after adopting folding this trend disappears.

Our results have been compared with the earlier results
obtained by Fraile et al.~\cite{Nicola}, Lang et al.~\cite{Weise}
and Mitra et al.~\cite{SSS}, which are attached in the upper and right 
panel of Fig.~(\ref{fig:eta_T_rs}).
Shear viscosity of pion gas, obtained by Refs.~\cite{Prakash,Gavin,SSS} 
in the kinetic theory approach, has been found as a monotonically increasing function
of temperature in the hadronic temperature range (0.100 GeV $< T <$ 0.175 GeV) for
vanishing baryon chemical potential ($\mu=0$) whereas Lang et al.~\cite{Weise} 
in the Kubo approach has shown $\eta$ as a decreasing function of temperature.
Similar decreasing nature of $\eta(T)$ are followed in the Kubo-approach calculation by 
Fraile et al.~\cite{Nicola} before unitarization of pion thermal width
but it turns to an increasing function
after dynamically generating low mass resonances ($\rho$ and $\sigma$)
via unitarization technique.
This change in nature of $\eta$ vs $T$ may be because of the transformation 
from the scenario of pionic medium without resonances 
to one with resonances.
Using the effective Lagrangian density, we have approximately mapped 
a similar kind of scenario, where probabilities of pion scattering 
with the low mass resonances $\rho$ and $\sigma$ have been found 
from the finite temperature calculation of pion self-energy for $\pi\rho$ and
$\pi\sigma$ loops.
Similar to the results obtained by Fraile et al.~\cite{Nicola} after unitarization, 
we have observed an increasing $\eta(T)$ after $T=0.100$ GeV 
but lower in magnitude and increasing slope.

Using ideal expression of entropy density $s$ for pion, we have shown
the variation of the ratio $\eta/s$ with respect to $T$ in the lower panel of
Fig.~(\ref{fig:eta_T_rs}), situated in the right side.
Within the hadronic temperature domain, 
our results of $\eta/s$ respect the KSS bound $\frac{1}{4\pi}$
as shown by horizontal red line. Since Niemi et al.~\cite{Niemi} have taken 
a similar range of $\eta/s(T)$, close to the KSS bound 
and found its important role
to fit $v_2(P_T)$ of RHIC data, therefore our estimation 
has a good association with phenomenological side also.

%
\section{Summary and Perspectives}
\label{sec:concl}
 
By considering the interaction of pion with the low mass resonances
$\sigma$ and $\rho$, we have calculated the shear viscosity coefficient 
of hot pion gas. In the framework of thermal field theoretical technique,
the thermal width $\Gamma_\pi$ has been calculated from one-loop pion self-energy
at finite temperature.
For the interaction part, we have taken an effective Lagrangian density,
by which pion self-energy for $\pi\sigma$ and $\pi\rho$ loops have been
evaluated at finite temperature. To treat $\sigma$ and $\rho$ 
resonances as two-pion states, we have folded the self-energy 
of $\pi\sigma$ and $\pi\rho$ loops
by the Breit-Wigner type spectral function of $\sigma$ and $\rho$ with their
vacuum width in $\pi\pi$ channel respectively. 
We have seen a complementary role of $\pi\sigma$ and
$\pi\rho$ loops to make $\eta$ be non-divergent
in the lower ($T<0.100$ GeV) and higher ($T>0.100$ GeV) temperature regions 
respectively. 
From the investigations of Niemi et al.~\cite{Niemi}, we see that $v_2(P_T)$ of RHIC data 
prefers these small values of $\eta/s(T)$ for hadronic matter.
This provides experimental justification to the microscopic calculation 
of shear viscosity due to
$\sigma\pi\pi$ and $\rho\pi\pi$ interaction performed in this work.

\end{document}